\documentclass[aps,nofootinbib,superscriptaddress,onecolumn,notitlepage,11pt]{revtex4-2}

\usepackage{amssymb, amsmath, amsthm, amsfonts} 
\usepackage{mathrsfs}
\usepackage{physics} 
\usepackage{mathtools}
\usepackage{hyperref} 
\usepackage{enumitem}
\usepackage{bm} 
\usepackage{csquotes} 
\usepackage{svg}
\usepackage{subcaption}

\numberwithin{equation}{subsection}

\usepackage[UKenglish]{isodate}
\usepackage[UKenglish]{babel}

\DeclareMathOperator{\E}{\mathbb{E}}
\DeclareMathOperator*{\argmax}{\mathrm{arg\,max}}

\newcommand{\lag}{\mathcal{L}}
\newcommand{\R}{\mathbf{R}}

\DeclareRobustCommand{\SkipTocEntry}[5]{}

\begin{document}

\title{A framework for the use of generative modelling in non-equilibrium statistical mechanics}

\author{Karl J Friston}
\affiliation{Queen Square Institute of Neurology, University College London, London, UK}
\author{Maxwell J D Ramstead}
\affiliation{Queen Square Institute of Neurology, University College London, London, UK}
\author{Dalton A R Sakthivadivel}
\email{dsakthivadivel@gc.cuny.edu}
\affiliation{Department of Mathematics, CUNY Graduate Center, New York, NY 10016, USA}

\date{\today}


\begin{abstract}

We discuss an approach to mathematically modelling systems made of objects that are coupled together, using generative models of the dependence relationships between states (or trajectories) of the things comprising such systems. This broad class includes open or non-equilibrium systems and is especially relevant to self-organising systems. The ensuing variational free energy principle (FEP) has certain advantages over using random dynamical systems explicitly, notably, by being more tractable and offering a parsimonious explanation of why the joint system evolves in the way that it does, based on the properties of the coupling between system components. The FEP is a method whose use allows us to build a model of the dynamics of an object as if it were a process of variational inference, because variational free energy (or surprisal) is a Lyapunov function for its dynamics. In short, we argue that using generative models to represent and track relations amongst subsystems leads us to a particular statistical theory of interacting systems. Conversely, this theory enables us to construct nested models that respect the known relations amongst subsystems. We point out that the fact that a physical object conforms to the FEP does not necessarily imply that this object performs inference in the literal sense; rather, it is a useful explanatory fiction which replaces the `explicit' dynamics of the object with an `implicit' flow on free energy gradients\textemdash a fiction that may or may not be entertained by the object itself. 

\end{abstract}

\maketitle


\tableofcontents 

\addtocontents{toc}{\SkipTocEntry}

\subsection*{Acknowledgements}

This paper more fully develops the theory outlined in a previous manuscript \cite{mtff}, and we are grateful to those listed there for their contributions to preliminary discussions on the topic. The account here is largely based on \href{https://darsakthi.github.io/talks/css-24}{a series of talks} given by the third named author at the Department of Philosophy of the University of Vienna, the Laufer Center for Physical and Quantitative Biology, the 2024 Conference on Complex Systems, and the informal discussions that followed. We thank the organisers and audiences of both. DARS acknowledges support from the Einstein Chair programme at the Graduate Centre of the City University of New York. KJF and DARS acknowledge support from the VERSES Research Lab.

\addtocontents{toc}{\SkipTocEntry}

\subsection*{Funding declaration}

There is no specific funding to be acknowledged. 

\addtocontents{toc}{\SkipTocEntry}

\subsection*{Author contributions}

The list of authors has been ordered alphabetically by surname. The CRediT contributions are as follows: KJF (formal analysis, software, writing---review and editing), MJDR (conceptualisation, writing---review and editing), DARS (conceptualisation, formal analysis, software, visualisation, writing---original draft).

\section{Introduction}


In statistics, generative models are joint probability distributions that model the relationships between different variables\textemdash most often, observations and their causes, or data and a set of labels. In the study of open and interacting systems, the use of statistical models\textemdash capable of describing such relationships\textemdash seems to be particularly apt. One approach to making such models, using variational (approximate Bayesian) inference \cite{Friston2019, ramstead2022bayesian, friston2023path}, describes how a physical object reflects the properties of its environment and {\it vice versa}. In this paper, we argue that generative models provide an especially useful foundation in scientific modelling generally, and especially in the study of systems exhibiting adaptation, morphogenesis, and other non-equilibrium phenomena. More specifically, we argue that generative modelling\textemdash premised on the use of variational inference to model the causes of the signals that a coupled system is receiving \cite{beal2003variational}\textemdash is particularly apt for the practice of modelling the activity of modelling itself; for instance, in modelling sophisticated self-organising systems, where we write down a representation of the model that we assume is in play in prediction and action. We use the ensuing variational free energy principle to illustrate how implementing models which (i) formalise how one object keeps track of another, and (ii) codify these relationships, can increase the fidelity of our mathematical modelling. 

Our proposed free energy principle (FEP) has been explored in some depth in prior literature, from the viewpoint of modelling control systems using the property that they generally look as if they track or infer the statistics of perturbations to their setpoints. In this paper we will make a more general case for its utility in statistical mechanics at large. After reviewing prior work on the FEP, we identify a particular issue that the FEP is able to uniquely address: the FEP keeps track of how a model depends on aspects of the thing that is modelled in a particularly insightful way, offering advantages in terms of tractability and explanation, by using gradients of variational free energy to describe the dynamics of interacting systems. 

One of the points this paper will emphasise is that using this framework does {\it not} necessarily presuppose that a system actually performs inference on its environment. It is a truism that physical systems need not explicitly calculate their trajectories of motion, to be modelled as pursuing such trajectories. We need not assume that the particle itself is literally performing inference when we build models under our framework. What is at stake is an `as if' description: we simply assume that there exists a quantity that varies systematically with the dynamics of the system; it so happens that \textit{surprisal} (or, equivalently, variational free energy) is a good choice of quantity for physical reasons, and that the minimisation of this quantity is mathematically equivalent to inference, in the sense of being an estimator. The fact that it is our model of the system which does inference and not necessarily the system itself is both a boon and a bane. It enables us to model the system using the more tractable gradient of variational free energy and the mathematics of statistical inference, rather than studying the possibly highly non-linear coupling between the dynamics of the two systems. However, it is not necessarily a faithful representation of the system in the literalist sense. (See also \cite{Andrews2021}.)

To that end, after describing the advantages of this FEP we will then argue that there is no conflation of the modelled system and our scientific model of it\textemdash in fact, we describe how they are carefully distinguished. This argument gives way to a more general theme of the FEP, which is that capturing interactions and dependence relations between things using generative models is insightful. We discuss how using the FEP gives us a flexible model of the properties of the object that look like they perform inference, but crucially, without inherently reifying the inferential aspects of the model, attributing them to the real-world object. That is to say, a model of an object can always be {\it written} as if it were modelling or inferring the structure of its environment. To make this argument, we appeal to two distinctions: one between the generative model (which is our scientific model) and the physical object being modelled (the real `thing' in the world), and the other, between the generative model and the variational density.

The structure of the paper is as follows. We first present a conceptual introduction to the FEP\textemdash and the way it leverages the relationships encoded by a generative model to build a {\it scientific model} of the dynamics of physical objects using free energy gradients. This is possible by modelling the property of physical systems that they \textit{look as if they engage in modelling} to an external observer. To sketch the argument: the FEP provides us with tools from dynamical systems and information theory, allowing us to model any coupled random dynamical systems as \textit{statistical estimators}, {\it i.e.}, as engaging in the statistical estimation of some quantity pertaining to the set of objects with which it interacts. Under the FEP, one object that is coupled to any other can thus be read as a representation of the coupled or estimated object, albeit of a deflationary, qualified sort (discussed in \cite{Ramstead2020semantics}; see also \cite{Ramstead2019enactive, ramstead2022bayesian}). More formally, the FEP applies to generative models that contain a partition of the states or paths of a joint system (a so-called `particular' partition, {\it i.e.}, into objects or particles). This partition induces a statistical boundary called a Markov blanket \cite{Pearl1998}, which captures conditional independence relations amongst subsets of the system \cite{Friston2019, sakthivadivel2022d}. The FEP says that, when such Markov blanketed subsets of the system exist, those subsystems track each others' statistical properties (dynamically) across the blanket \cite{classical-physics}. Essentially, the FEP is a modelling framework which captures the following fact: models of coupled particles or factorised systems can be written as parameter estimators, and estimating the parameters of a distribution is a form of Bayesian inference. 

We then argue that there is no ambiguity between model and modelled real-world system in this formalism. We can think of the FEP, metaphorically, as a map of the part of the territory that \textit{behaves as if it were a map}\textemdash or looks as such to an external observer. Accounts such as \cite{Andrews2021} have encouraged practitioners of the FEP to be explicit about this point. Viewing the theory this way, not only is there no conflation between our FEP-theoretic model of the system and the system itself, but in fact we arrive at a novel modelling method for coupled random dynamical systems. In particular, we do not claim that the objects in the system themselves must literally be performing inference. Rather, we describe the dynamics as doing inference, in the sense of being a statistical estimator whose states parameterise a variational density.

In summary we propose that the FEP provides a map of any possible map whatsoever\textemdash of, or held by\textemdash a physical system. (This includes the use of the FEP itself by a modeller.) We suggest that this leads to constraints on all possible maps, or models of physical things, arising from what it means to be a map or model of a physical process at all. In the fashion of Jaynes, who proposed that doing statistical mechanics is doing inference about the probabilities of microstates given knowledge of macrostates \cite{jaynes1, jaynes2}, the FEP is a natural expression of statistical physics as a process of making models. More broadly we claim that this echoes the sentiments of Wittgenstein, providing a constraint on modelling in terms of logically necessary preconditions for sense-making and the parsing of sensory streams \cite{Ramstead2020semantics}.

For those already familiar with the FEP, this paper can be thought of as going back to basics, emphasising its foundation in inverting generative models (written as stochastic differential equations (SDEs) in generalised coordinates), using a path integral scheme for minimising variational free energy \cite{friston2010generalised}.

\section{Summary remarks on the free energy principle}\label{summary-section}

\subsection{Fundamentals}


Here we will describe the proposed free energy principle (also directing the reader to prior literature such as \cite{classical-physics, ramstead2022bayesian, friston2023path} for a more extended look). The FEP aims to describe the physical properties of a system in terms of dynamics on a manifold of probability distributions ({\it i.e.}, as variational inference) \cite{DaCosta2021}, resting on the useful property that physical objects generally look as if they track or infer the statistics of things to which they are coupled \cite{ramstead2022bayesian, Sakthivadivel2022b, classical-physics}. Formulated most generally, the FEP says that the average path taken by an object that is sparsely coupled to its environment is characterised by a minimisation of a variational free energy functional. 

We consider a particle (also referred to as a `system') with states $\mu$ coupled to another system with states $\eta$ (henceforth, the `environment') through a set of variables $B$ with values $b$, and a steady state density $p(\eta, b, \mu)$ such that $\mu$ is conditionally independent of $\eta$ given $B$. After Pearl (see \cite{Pearl1998}) we refer to $B$ as the {\it Markov blanket} of the system. Typically, $B$ will itself consist of two different kinds of states---sensory states, providing the system signals about its environment, and active states, providing interventions of the system on the environment. The coupling at play puts us in a non-equilibrium situation, where a system can be driven by fluxes of energy or matter from other systems in its environment \cite{seifert2012stochastic}. 

The FEP begins by saying that if there exists a boundary in the form of a Markov blanket, then there exists a function $\sigma$ relating (the dynamics of) internal states to the environment across the boundary, and in particular, such that $\sigma(\hat\mu_b) = \hat\eta_b$, where $\hat{\,\cdot\,}_b$ denotes the conditional expectation given the blanket state $b$ \cite[Lemma 4.3]{Sakthivadivel2022b}. Following that, the aim of the FEP is to understand what the generative model $p(\eta, b, \mu)$ says about the dynamics of the subsystem with states $\mu$, beyond just its marginal distribution. In particular: if we assume that $\hat\eta_b$ parametrises $p(\eta \mid b)$, then we have made $\mu$ into an {\it estimator} of the statistics of the environment. 

The FEP is simplest when we have a non-equilibrium steady state density, in which case the following claim holds: if the system remains on some attractor of states, it looks like it minimises variational free energy. This is true in the following sense. It is possible to express the dynamics of a random variable with a non-equilibrium steady state density 
\[
    \dd{X_t} = f(X_t) \dd{t} + D(X_t)\dd{W_t}
\]
in terms of a gradient flow on the surprisal weighted by a particular matrix field \cite{ao2004potential, ma2015, yuan2017sde, da2023entropy}, having the form 
\begin{equation}\label{sd-eq}
    \dd{X_t} = -(Q(X_t) - \Gamma(X_t))\nabla_x \log p^*(X_t) \dd{t} + D(X_t)\dd{W_t}
\end{equation}
where $Q(x)$ is skew-symmetric everywhere, $\Gamma(x)$ is positive (semi-)definite everywhere and satisfies $2\Gamma(x) = D(x) D^{\top}(x)$, and $p^*(x)$ is the non-equilibrium steady state density. This decomposition holds whenever there exists a $Q(x)$ such that
\[
\nabla \times [Q(x) + \Gamma(x)]^{-1} f(x) = 0
\]
and $\Gamma$ is divergence free---otherwise, in the It\=o picture, a correction is necessary due to the divergence of $\Gamma$. For an equilibrium steady state, $Q(x)$ is identically zero and $f$ is the gradient of some scalar function on the nose; letting $U(x)$ be that scalar function, the steady state is then proportional to $\exp{-U(x)}$. 

Moreover, the log-probability is able to describe the probability of all fluctuations of the system in terms of their distance from the most likely state, as well as physically-derived quantities \cite{freidlin_wentzell_2012, seifert2012stochastic, seifert2019stochastic}, making this a meaningful equation. (See \cite{parr2020markov} where this is applied to reproduce certain aspects of stochastic thermodynamics.) 

Let us make the Laplace assumption such that the most likely state is the expectation. Then our generic SDE \eqref{sd-eq} tends towards the fixed point solution 
\[
    \hat\mu_b = \argmax \,\log p(\mu, b).
\]
To make use of this first observation, we now {\it postulate} a conditional density over external states, $q(\eta; \hat\eta_b)$, where we have parameterised the variational density by the conditional mode $\hat\eta_b$ of the environment. Under the Laplace assumption, in a neighbourhood of its peak the posterior is approximately Gaussian, and hence uniquely determined by its mean and variance. This guarantees that finding the right value of $\hat\eta_b$ is sufficient for $q$ to equal the posterior almost surely. More generally, under the maximum entropy assumption of a Gibbs density, any constraint function suffices. Once again, this implies that the system's average behaviour encodes an inference about its environment, in the sense of parameter estimation.

Inference by parameter estimation implies the Kullback--Leibler (KL) divergence between the posterior and the true density vanishes as the parameter varies. We will write
\[
    F(\mu, b) \coloneqq \int q(\eta; \sigma(\mu)) \log q(\eta; \sigma(\mu)) \dd{\eta} - \int q(\eta; \sigma(\mu)) \log p(\eta \mid b) \dd{\eta} - \log p(\mu, b).
\]
Now observe that since the divergence term is non-negative, $F(\mu, b)$ is an upper bound on the surprisal $-\log p(\mu, b)$,
\begin{equation}\label{abil-ineq}
    F(\mu, b) = D_{\mathrm{KL}}(q(\eta; \sigma(\mu)) \|\,  p(\eta \mid b)) - \log p(\mu, b) \geqslant - \log p(\mu, b)
\end{equation}
with equality precisely when the KL term vanishes. We conclude that the minimisation of surprisal implies the vanishing of the divergence term. Let us apply the Laplace approximation \cite{friston2007variational} such that the expected state is approximately the mode of the density. It follows that the most probable state is the least surprising one. Hence, for physical reasons, we expect the surprisal to be minimised. From that reasoning, \eqref{sd-eq} has the equivalent form
\begin{equation}\label{eq-of-motion}
    \dd{\mu_t} = -(Q - \Gamma) \nabla_\mu F(\mu, b) + \dd{W_t}.
\end{equation}
Evidently, when the inference gap is minimised, we have
\[
-(Q - \Gamma) \nabla_\mu \log p(\mu, b) + \dd{W_t} = -(Q - \Gamma) \nabla_\mu F(\mu, b) + \dd{W_t}
\]
on the nose. Our reasoning so far implies this suffices as an approximate description of all Markov blanketed systems. 

What we have done is used the fact that the dynamics of a (steady state) system minimise surprisal on average, and argued that whenever equipped with a Markov blanket they also minimise variational free energy; and hence, the dynamics of a system with a particular partition can always be written as a gradient flow on variational free energy. This gradient flow is what we identify as actually being the process ({\it i.e.}, the performance) of inference. Explicit examples of this can be found in \cite{parr2020markov, ueltzhoffer2021drive, classical-physics}. The particular partition becomes crucial here, since this is what we leverage in our model to incorporate some unknown quantity (external states beyond the blanket) that the system can do inferences about at all. Variational free energy, being a tractable ({\it i.e.}, easily computed) upper bound on surprisal, already makes this a useful tool to have---however we will go further and claim supposing the existence of a minimised variational free energy, with its interpretation as inferring a target density by approximating it with a variational density, has a use in interpreting models in statistical physics in a new way.

We will remark that much of this story can be told for the trajectories of a system and distributions over path spaces, but with many modifications \cite{classical-physics, friston2023path}, none of which are necessary to make the point sought in this paper.

\subsection{Analysis and interpretation}

We have, in the previous subsection, introduced a vanishing KL divergence term to our SDE-based model of a physical system, which we have then stipulated to vanish. Though seemingly vacuous, this move opens a world of interesting interpretations for models built according to this reasoning, by connecting the physically-given dynamics of the mode of the system to inference performed by the mode of an estimator. For example: by existing a particular way inside the joint state space, the particle can be seen as {\it encoding} ({\it via} its mode) an inference about a generative model over the entire joint state space, resulting from carving out some `niche' within it. That is to say, the environment must likely be some certain way for the system to exist the way it does. On the other hand, if free energy (or surprisal) remains irreducibly high, it is likely that the particle occupies surprising ({\it i.e.}, implausible or uncharacteristic) states; this would render it non-existent over some timescale, in the sense that it would be occupying states that were not characteristic of the thing in question ({\it e.g.}, a fish out of water). Correspondingly, its inference about the generative model would be incorrect, concentrating mass in the wrong areas of the joint state space. These statements tie together the reasoning that things which exist in an environment reflect the statistics of their environment in a particular way, based on the way they are coupled to their environment; or, they do not exist that way.

Here, variational free energy is a Lyapunov function\textemdash a function that decreases monotonically along the trajectories of a (deterministic) dynamical system. The utility of Lyapunov functions is that they guarantee stability, allowing us to understand the dynamical system as flowing on the gradient of the Lyapunov function. Variational free energy arises as a Lyapunov function for the straightforward reason that the most probable state is the least surprising one, and that the log-probability is able to describe the probability of all fluctuations of the system in terms of their distance from that most likely state \cite{freidlin_wentzell_2012, seifert2012stochastic, seifert2019stochastic}---combined with the inequality in \eqref{abil-ineq}. Since the dynamics towards the mode induce a gradient flow on a divergence between the estimated parametric distribution and the true distribution, the FEP moreover proposes that these dynamics (in virtue of the coupling) can be modelled as a variational inference process, following a gradient flow on the variational free energy. Said another way, the crucial idea is that the dynamics of a system do inference when written as minimising a free energy functional, which is itself the consequence of reading surprisal as a Lyapunov function for those dynamics, following {\it e.g.} Freidlin--Wentzell, and noting that the KL divergence is {\it stipulated} to be minimised by a near-tautology that the system will be at its average state on average. In that case, we can map these parameters to $\sigma(\hat\mu_b)$, meaning that for as long as the particle is at state $\hat\mu_b$, the \emph{particle} models its environment by estimating the parameters of its distribution. 

Ultimately this simply means that, in virtue of being coupled ({\it via} blanket states) to its environment, the particle reflects data about its environment\textemdash or, what it believes that its environment is like. Under the assumption that a particle which exists in an environment models its environment, free energy is implicitly a measurement of coherence. If the particle stops modelling its environment in a particular way ({\it i.e.}, for a particular $\sigma$), it must have ceased to exist with the mean that $\sigma$ was constructed for; conversely, if a particle ceases to exist, it will stop modelling its environment. The surprisal being considered is, much like probability, surprisal given a model, in that it quantifies how surprising it would be for an observer\textemdash equipped with a model\textemdash to find the system (such as itself) in some non-system-like state or configuration. Minimising surprisal implies that the variational free energy is minimised, ({\it i.e.}, the object-as-model is a good estimator of its environment \cite{classical-physics}), and likewise, minimising the variational free energy places an upper bound on the surprisal.

Similarly, note the {\it direction} of the implication in our claims. In the context of modelling persistent systems, the FEP is a sort of inversion of the typical reasoning one sees: instead of saying an object persists (as a thing) because it is stable, we say an object is stable (as what it is) because it persists (as that thing) \cite{Sakthivadivel2022b}. We {\it stipulate} that the KL divergence vanishes\textemdash because it is a tautology that $p$ should equal $q$ almost surely\textemdash and because physics only `works' in a given regime if this is the case (see \cite[\S IIB]{classical-physics} where this argument is spelt out). This happens in the background, so to speak. Then we can use the FEP to describe the dynamics as a gradient flow on surprisal\textemdash which can be done by a simple mathematical fact and independently of the FEP\textemdash as a process which {\it has performed} inference {\it in virtue of existing}. Adding this `inflationary' KL term, the non-equilibrium steady state density of the system can be understood as a model of the environment under this stipulation. All this is made possible by the existence of a particular partition. Hence, we arrive at a deflationary account of intelligence through an inflationary mathematical technique. However---note that when the Markov blanket consists of sensory states and active states, this inference can immediately be written as part of a control problem \cite{friston2012active, pezzulo2015active, pio2016active, pezzulo2018hierarchical, baltieri2019pid, millidge2020relationship, baioumy2021active}, lending it an even more sophisticated interpretation.

From the viewpoint of model building, since internal states are by construction unobservable, we can in principle tell any story that we like about them\textemdash provided that this story is supported well enough by observable evidence. One such story is that those internal states look as if they parameterise a variational density over external states. Taking this explanation seriously, and leveraging it as an approach to modelling physical systems, \textit{is} applying the FEP.

\section{Empirical validation}

In this section we bolster the claim that modelling physical systems with variational free energy is advantageous by reviewing some simulations of self-organisation and self-assembly. Two toy models will be studied---cellular morphogenesis, and an ensemble of excitable cells. In these simulations we will see that introducing the generative model allows us to model the system self-organising as if it `knew its place' within a larger system---inferring the unknowable external states by the minimisation of variational free energy.

\subsection{Cellular morphogenesis}

Starting from an ensemble of undifferentiated cells at the same location, we wanted to simulate their migration and differentiation using the minimisation of variational free energy. The simulation here exploits the fact one can express a Markov blanketed system's marginal Lyapunov function in terms of a variational free energy function, such that its attracting set can be prescribed in terms of the generative model that defines the free energy. In essence, the system's dynamics rest upon each cell inferring its unique identity (in relation to all others) and behaving in accord with those inferences.\footnote{These simulations were based on \cite{friston2015knowing} and use the DEM toolbox in the MATLAB software SPM12 (\url{https://www.fil.ion.ucl.ac.uk/spm/}). They can be reproduced using the morphogenesis feature and the data provided there.}

A group of eight cells was simulated with states consisting of the physical locations on a two-dimensional grid ($\psi_x$) of each cell and the cell type of each cell (measured by its cell signalling profile, a four-dimensional vector of signals expressed $\psi_c$). For cell $i$, $(\psi_{xi}, \psi_{ci})$ is an internal state $\mu$, and the remaining tuples $(\psi_{xj}, \psi_{cj})$ constitute external states $\eta$. 

Each cell was equipped with active states $(a_x, a_c)$ corresponding to modulations of its position and chemotactic signalling profile, respectively. We assume the active states fluctuate quickly enough that from the viewpoint of the environment their dynamics are negligible, {\it i.e.} $\psi \approx a$, constituting an adiabatic approximation
\[
    \tau \partial_t \psi = a - \psi
\]
with large time constant $\tau$. In this way the cell's active states become the external states offered to other cells. (Note that this assumption is purely for convenience, simplifying the analytic expressions for certain quantities; it is by no means necessary in the following discussion, and the example to follow will not use it. What is at stake in terms of the realism of the simulation is: the adiabatic assumption excludes the possibility of modelling cell migration or signal expression as a function of action.)

We will additionally assume an exponential decay in space of extracellular signal concentration as a function of the cell's own signalling. This means the cell can sense 
\[
    \lambda_i(\psi_{x}, \psi_{c}) = C\sum_j \psi_{cj} \,e^{\abs{\psi_{xi} - \psi_{xj}}}, \quad C \in [0,1].
\]
Each cell also has sensory states detecting the cell's current location and the concentration of chemical signals produced by the cell. We will assume these are simply noisy estimates of $(\psi_{xi}, \psi_{ci})$ with noise distributed like a Gaussian of precision $\Pi^{(1)} = e^{-16}$, for the full sensory state
\[
    s_i = [\psi_{xi} + \omega, \psi_{ci} + \omega, \lambda_i(\psi_x, \psi_c) + \omega].
\]
In summary, the active states are (i) where a cell chooses to be and (ii) what a cell chooses to release. Those actions are the external states being offered to other cells under the adiabatic assumption, and through the kernel above they superpose into a shared field which every cell samples.

A point in the state space $(\psi_x^*, \psi_c^*)$ was designated as a target morphology, giving prescribed locations and cell types of all cells. This preferred morphology was encoded in a prior and specified desired positions and signal expressions for each clone or cell in terms of the cell's expectations regarding sensory signals and spatial configurations. This point consisted of the physical form of a head, body, and tail, and with different cell types in each component of the organism. The target sensation under $\lambda$ is plotted in Figure \ref{main:a}. In real situations one can imagine a genetic encoding of the target morphology, with epigenetic dynamics being active inference. A possible example of this is given in Figure \ref{main:b} where the target is encoded in binary codons for location and expression. We then equipped each cell with a generative model, specifying it as follows. The mean was given in terms of how environmental states are mapped to sensory states, 
\[
    g(\psi)_i = \frac{\exp(\psi_i)}{\sum_j \exp(\psi_j)}
    \left[
    \begin{array}{c}
        \psi_x^*\\
        \psi_c^*\\
        \lambda(\psi_x^*, \psi_c^*)\\
    \end{array}
    \right]
\]
with the softmax function returning the expected identity of each cell. In Figure \ref{main:c} we show the softmax function, interpreted as the posterior beliefs that each cell (column) occupies a particular place in the ensemble (rows). A precision $\Pi^{(2)} = e^{-2}$ was also set over external states. A {\it prediction error} for how the generative model matched with sensations was given in the form 
\[
    \varepsilon_i = [s_i - g(\psi)_i]
\]

The model was operationalised by following the procedure in \cite{friston2010generalised}. A Taylor expansion of trajectories was taken and the statistics of each order of approximation $(\psi, \psi', \psi'', \ldots)$ were modelled at each point in time. A derivative operator $D$ was defined for ordinary differential equations on this space. The entire trajectory was given a coordinate in function space
\[
    \tilde\psi = [\psi, \psi', \psi'', \ldots].
\]
Now by the decomposition \eqref{sd-eq}, we have the equations of motion
\begin{equation}\label{ex-eq-of-motion}
    \begin{split}
        \dd{\tilde{\mu}_t} &= -(Q_{\mu\mu} - \Gamma_{\mu\mu})\nabla_{\tilde \mu} \log p^*(\tilde s, \tilde a, \tilde \mu)\dd{t} + \dd{\tilde{W}^\mu_t},\\
        \dd{\tilde{a}_t} &= -(Q_{aa} - \Gamma_{aa})\nabla_{\tilde a} \log p^*(\tilde s, \tilde a, \tilde \mu)\dd{t} + \dd{\tilde{W}^a_t}
    \end{split}
\end{equation}
for each cell.\footnote{This follows from the marginal flow lemma of \cite{Friston2019}.} This is where the interpretation as inference which we supply as modellers comes into play. We cannot compute the solution to \eqref{ex-eq-of-motion} since the surprisal term yields an intractable integral. However, under the generative model we have equipped our system with, we can instead treat the dynamics of the cell under \eqref{ex-eq-of-motion} as maximising log-evidence for the sensations of the cell, and hence as a variational problem---as posed throughout \S\ref{summary-section}. We will introduce the variational distribution 
\[
    q(\tilde \psi \mid \tilde \mu) = \mathcal{N}\big(\tilde\mu; -\nabla_{\tilde \mu \tilde \mu} \log p(\tilde \mu, \tilde s, \tilde a, \tilde \mu)\big), 
\]
a Gaussian with expectation being the coordinate of the internal trajectory in function space $\tilde \mu$ and variance being the curvature of the surprisal---computed at $\tilde\psi = \tilde \mu$---and free energy
\[
F(\tilde a, \tilde s, \tilde \psi) = - \log p(\tilde s, \tilde a, \tilde \mu) + D_{\mathrm{KL}}\big(q(\tilde\psi \mid \tilde \mu) \,\|\, p(\tilde\psi \mid \tilde s, \tilde a, \tilde \mu)\big). 
\]
Treating \eqref{ex-eq-of-motion} as a Bayesian filter of trajectories, from the derivation outlined in \cite[equation 3.6]{friston2010generalised} the final drift terms for the equations of motion for the $i$-th cell given the generative model were obtained as (again denoting the internal state $\psi_i$ as $\mu$)
\begin{equation}\label{example-eq}
    \begin{split}
        f_{\tilde\mu}(\tilde a_i, \tilde s_i, \tilde \mu) &= (Q_{\mu\mu} - \Gamma_{\mu\mu})\nabla_{\tilde\mu} F(\tilde a_i, \tilde s_i, \tilde \mu) = D\tilde\mu - \nabla_{\tilde\mu}\tilde\varepsilon_i \cdot \Pi^{(1)}\tilde\varepsilon_i - \Pi^{(2)}\tilde\mu \\
        f_{\tilde a_i}(\tilde a_i, \tilde s_i, \tilde \mu) &= (Q_{a_i a_i} - \Gamma_{a_i a_i})\nabla_{\tilde a_i} F(\tilde a_i, \tilde s_i, \tilde \mu)  = D\tilde a_i - \nabla_{\tilde a_i} \tilde s_i \cdot \Pi^{(1)}\tilde\varepsilon_i
    \end{split}
\end{equation}
with the following explicit equations for (positional and signalling, respectively) active states
\begin{align*}
    \partial_t \tilde a_x &= D\tilde a_x - \nabla_x \tilde s_x\cdot \Pi^{(1)}_x \tilde\varepsilon_x + \nabla_x \tilde s_\lambda \cdot \Pi^{(1)}_\lambda \tilde\varepsilon_\lambda\\
    \partial_t \tilde a_c &= D \tilde a_c - \Pi^{(1)}_c \tilde\varepsilon_c + \Pi^{(1)}_\lambda \tilde\varepsilon_\lambda.
\end{align*}
As before, if the inference gap is minimised, for instance, in the neighbourhood of the mode, then \eqref{example-eq} is equal to \eqref{ex-eq-of-motion} on the nose. 

We then simulated the dynamics of each cell. In Figure \ref{figure2} we have shown the expected positions and releases of each cell over time, extracted from the solution to \eqref{example-eq}, and the decrease in free energy as the cells evolve towards the target state. 

In summary, the cell carries a generative model that predicts what signals it should sense if it were the `correct' type of cell at the `correct' place in the target morphology. Those predictions are built from the target locations and target signal combinations, and the model maps internal beliefs about identity through the generative model to predicted sensory consequences. The final outcome of this process is shown in Figure \ref{final-loc-fig}. 

\begin{figure}[htb!]

\centering
\subfloat[]{\label{main:a}\includesvg[scale=0.6]{target.svg}}

\par\medskip

\begin{minipage}{.5\linewidth}
\centering
\subfloat[]{\label{main:b}\includesvg[scale=0.6]{encoding.svg}}
\end{minipage}%
\begin{minipage}{.5\linewidth}
\centering
\subfloat[]{\label{main:c}\includesvg[scale=0.6]{softmax.svg}}
\end{minipage}

\caption{Plots of the target extracellular gradients (\ref{main:a}), encoding of the target signal in the cells (\ref{main:b}) and softmax expectations of the identities of each cell (\ref{main:c}). In Figure \ref{main:a} colours denote different signal expressions and black dots denote no data, with the final locations of the cells starred.}
\label{fig:main}
\end{figure}

\begin{figure}[htb!]
    \centering
    \begin{subfigure}[t]{0.5\textwidth}
        \centering
        \def\svgscale{0.8}
        \includesvg{cell-migration.svg}
        \caption{{\bf Migration of cells in time.} As the cells differentiate they are marked with red, yellow, green, or blue, according to the signal they follow, with the gradient denoting the level of how strongly a signal is expressed.}
        \label{migration-fig}
    \end{subfigure}%
    ~
    \begin{subfigure}[t]{0.5\textwidth}
        \centering
        \def\svgscale{1}
        \includesvg{free-energy.svg}
        \caption{{\bf Decrease of free energy in time}. A decrease of free energy in time is noted as the cells migrate to the final states corresponding to the sensed target signal.}
    \end{subfigure}%
    \caption{Here morphogenesis is demonstrated by cells migrating to align with expectations based on sensory input, modelled by a concomitant decrease of free energy in time.}
    \label{figure2}
\end{figure}

\begin{figure}[htb!]
    \centering
    \includesvg{extrinsic2.svg}
    \caption{{\bf Dynamics of each cell, with final locations starred.} The gradients here match those of Figure \ref{migration-fig}. It can be seen that (speaking somewhat heuristically) each cell moves to fulfil its {\it expectations} about the signals it should encounter, whilst expressing the signals associated with its current beliefs about its place in the target ensemble.}
    \label{final-loc-fig}
\end{figure}

\subsection{Periodically-firing cells}

A similar but more complicated example is a ring of excitable, gap-junction-coupled cells whose generative model encodes a periodic target waveform. Each cell's internal states infer the current phase of that cycle from its membrane sensors, whilst active states like ion-channel gating act to make sensed signals match the predicted periodic trajectory. This model is more expressive because the prior over hidden states is oscillatory, such that the free energy minimum is a limit cycle: the population converges to a stable circulating orbit in state space rather than to a fixed point.

We will define external $\eta_t$ as a stimulus in the world, $s_t$ as a sensory state, $\mu_t$ as the internal state estimating the phase of the cell, and $a_t$ as the state of some actuator or motor acting on the environment to make observations conform to predictions. For simplicity we will assume $s_t$ tracks $\eta_t$ and that sensor noise is standard white in time, {\it i.e.} that the likelihood satisfies
\[
    p(s \mid \eta) \propto \exp{-\frac{(s - \eta)^2}{2\sigma_s^2}}
\]
at steady state. We will also introduce a likelihood over outcomes of actions
\[
    p(\eta \mid a) \propto \exp{-\frac{(\eta - a)^2}{2\sigma_\eta^2}}
\]
in place of the simplifying adiabatic assumption used in the previous example. 

The system's predictions of sensations given a phase follow the cosine function $s(\mu) = A \cos \mu$ and we introduce a predictive control prior making actions prefer to create the encoded preference for the waveform,
\[
    p(a \mid \mu) \propto \exp{-\frac{(a - A \cos\mu)^2}{2\sigma_a^2}}.
\]
There will finally be a uniform prior on $\mu$ as a circle-valued variable. 

In total we have a one-step generative model (conditioned on the internal phase):
\[
    s \mid \eta \sim \mathcal{N}(\eta,\sigma_s^2), \qquad
    \eta \mid a \sim \mathcal{N}(a,\sigma_\eta^2), \qquad
    a \mid \mu \sim \mathcal{N}(A\cos \mu,\sigma_a^2).
\]
The generative model is linear in $(s, \eta, a)$ and mildly nonlinear in $\mu$ through the cosine function. To keep this example brief---and to make the surprisal analytically soluble, such that some direct comparisons can be made between \eqref{sd-eq} and \eqref{eq-of-motion}---we will not create any hidden states to make inferences about beyond the environmental state, so our generative model is not hierarchical and generalised coordinates of motion need not be used. 

In that case, using conditional independence, we can write at steady state (up to additive constants)
\[
    - \log p(x) = \frac{(s - \eta)^2}{2\sigma_s^2} + \frac{(\eta - a)^2}{2\sigma_\eta^2} + \frac{(a - A \cos\mu)^2}{2\sigma_a^2}.
\]
Now specify a parametrisation of $Q$ and $\Gamma$,
\[
    Q = 
    \begin{bmatrix}
        0 & q_{\mu s} & q_{\mu a} & 0\\
        -q_{\mu s} & 0 & 0 & 0 \\
        - q_{\mu a} & 0 & 0 & q_{a \eta}\\
        0 & 0 & -q_{a \eta} & 0
    \end{bmatrix}, 
    \qquad \qquad \Gamma = \mathrm{diag}(\gamma_\mu, \gamma_s, \gamma_a, \gamma_\eta).
\]
Notice there are no couplings in the flow over the state space between $\mu$ and $\eta$. This is an indication of the Markov blanket. 

Since the generative model is somewhat simple, it is straightforward to write the equations of motion:
\begin{equation}\label{example-eq-ii}
    \begin{split}
        \partial_t \mu_t &= q_{\mu s} \partial_s L + q_{\mu a} \partial_a L - \gamma_\mu \partial_\mu L + \xi_t^\mu \\ &= q_{\mu s} \frac{s - \eta}{\sigma_s^2} + q_{\mu a} \left(\frac{a-\eta}{\sigma_\eta^2} + \frac{a - A \cos \mu}{\sigma_a^2}\right) - \gamma_\mu \frac{A \sin \mu}{\sigma^2_a} (a - A \cos \mu) + \xi_t^\mu \\
        \partial_t s_t &= -q_{\mu s} \partial_\mu L - \gamma_s \partial_s L + \xi_t^s \\
        &= -q_{\mu s}  \frac{A \sin \mu}{\sigma^2_a} (a - A \cos \mu) - \gamma_s \frac{s - \eta}{\sigma_s^2} + \xi_t^s \\
        \partial_t a_t &= -q_{\mu a} \partial_\mu L + q_{a\eta} \partial_\eta L - \gamma_a \partial_a L + \xi_t^a \\
        &= -q_{\mu a} \frac{A \sin \mu}{\sigma^2_a} (a - A \cos \mu)  + q_{a\eta} \left( \frac{\eta - s}{\sigma^2_s} + \frac{\eta - a}{\sigma_\eta^2} \right) - \gamma_a \left(\frac{a-\eta}{\sigma_\eta^2} + \frac{a - A \cos \mu}{\sigma_a^2}\right) + \xi_t^a \\
        \partial_t \eta_t &= -q_{a\eta} \partial_a L - \gamma_\eta \partial_\eta L + \xi_t^\eta\\
        &= -q_{a\eta}\left(\frac{a-\eta}{\sigma_\eta^2} + \frac{a - A \cos \mu}{\sigma_a^2}\right) - \gamma_\eta \left( \frac{\eta - s}{\sigma^2_s} + \frac{\eta - a}{\sigma_\eta^2} \right) + \xi_t^\eta
    \end{split}
\end{equation}
where $\xi^i_t$ are independent Brownian motions. Notice the solenoidal couplings between sensory states and environmental states preclude any adiabatic assumption. 


Using properties of marginals of jointly Gaussian random variables we can compute the steady state 
\[
    p^*(s, a, \mu) \propto \int e^{-L(\mu,s,a,\eta)} \dd{\eta} = \frac{1}{2\pi} \mathcal N\big(a; A\cos \mu, \sigma_a^2\big) \mathcal N\big(s; a, \sigma_s^2+\sigma_\eta^2\big)
\]
{\it i.e.} i.e. a uniform phase prior, a Gaussian action prior around $A \cos \mu$, and a sensor likelihood concentrated around the action with variance $\sigma_s^2+\sigma_\eta^2$. The interpretation of this density is that the triple $(\mu, s, a)$ forms a ribbon-shaped non-equilibrium steady state in $S^1 \times \R^2$ aligned with the subspace $\{a = A \cos \mu, s = a\}$. The solenoidal component makes the probability circulate along this ribbon; however, the shape of the ribbon is fixed by the generative terms specified above. 

The system can be treated as a linear Gaussian chain, with Gaussian likelihoods for $s \mid \eta$ and $\eta \mid s, a$, providing an exact expression for the posterior
\[
    p(\eta \mid s, a) \propto \exp{- \frac{(s - \eta)^2}{2\sigma_s^2} - \frac{(\eta - a)^2}{2\sigma_\eta^2}}.
\]
Completing the square in $\eta$ yields $p(\eta \mid s, a) = \mathcal{N}(m_{\mathrm{post}}, v_{\mathrm{post}})$ where 
\[
    v_{\mathrm{post}} = \left(\frac{1}{\sigma_s^{2}} + \frac{1}{\sigma_\eta^{2}}\right)^{-1}, \qquad m_{\mathrm{post}} = v_{\mathrm{post}}\left( \frac{s}{\sigma^2_s} + \frac{a}{\sigma^2_\eta}\right)
\]
so that the posterior is parametrised by a variance of twice the inverse precision. Now let the internal states parametrise a Gaussian posterior expressing beliefs about the world, $q(\eta; \mu, \tau^2) = \mathcal{N}(\eta; A \cos \mu, \tau^2)$.  The free energy can be expressed analytically as
\begin{align*}
    F(s, a, \mu) &= \E_q[\log q(\eta; \mu, \tau^2)] - \E_q[\log p(\eta \mid s, a)] - \log p(s, a, \mu) \\ &= -\frac{1}{2} \log 2\pi \tau^2 - \frac{1}{2} + \frac{1}{2}\log 2\pi v_{\mathrm{post}} + \frac{1}{2 v_{\mathrm{post}}} \left( \tau^2 + (A \cos \mu - m_{\mathrm{post}})^2\right) - \log p(s, a, \mu)
\end{align*}
or, collecting like terms, 
\[
\frac{1}{2}\left( \log \frac{v_{\mathrm{post}}}{\tau^2} + \frac{\tau^2 + (A \cos \mu - m_{\mathrm{post}})^2}{v_{\mathrm{post}}} - 1 \right) - \log p(s, a, \mu)
\]
so that when $\tau^2 = v_{\mathrm{post}}$ and $m_{\mathrm{post}} = A \cos \mu$, the KL term vanishes. 

Taking gradients of the free energy yields the equations of motion 
\begin{equation}\label{example-eq-iii}
    \begin{split}
        \partial_t \mu_t &= q_{\mu s} \partial_s F + q_{\mu a} \partial_a F - \gamma_\mu \partial_\mu F + \xi_t^\mu \\ 
        &= q_{\mu s} \frac{s - A \cos \mu}{\sigma_s^2} + q_{\mu a}(a-A \cos \mu)\left(\frac{1}{\sigma_\eta^2} + \frac{1}{\sigma_a^2}\right) \\ 
        &\phantom{--} - \gamma_\mu A \sin \mu \left( \frac{s - A \cos \mu}{\sigma_s^2} + \frac{a - A \cos \mu}{\sigma^2_\eta} + \frac{a - A \cos \mu}{\sigma_a^2} \right) + \xi_t^\mu \\
        \partial_t a_t &= -q_{\mu a} \partial_\mu F - \gamma_a \partial_a F + \xi_t^a \\
        &= -q_{\mu a} A \sin \mu \left( \frac{s - A \cos \mu}{\sigma_s^2} + \frac{a - A \cos \mu}{\sigma^2_\eta} + \frac{a - A \cos \mu}{\sigma_a^2} \right) \\
        & \phantom{--} - \gamma_a (a-A \cos \mu)\left(\frac{1}{\sigma_\eta^2} + \frac{1}{\sigma_a^2}\right) + \xi_t^a.
    \end{split}
\end{equation}
We then simulated the system as a Kalman update for a linear Gaussian state space channel $\eta \to s$ with input $a$.\footnote{The python code for this simulation is available at \url{https://github.com/DARSakthi/periodically_firing_cells/}.} In practice, this means when simulating the system forwards, we set the parameters of the variational density to those of the posterior such that the variational density $q$ became the (instantaneous) Kalman posterior over the latent world state $\eta$ and the instantaneous precision-weighted prediction errors drove the gradient descent on $F$.

In Figure \ref{fig:internal} sampled trajectories of $\mu_t$ under \eqref{example-eq-ii} and \eqref{example-eq-iii} are compared (Figure \ref{fig:actions}, $a_t$, respectively). Close correlation is observed between the two flow regimes. A corresponding tendency towards zero of the sample mean of the KL term is observed in Figure \ref{fig:KL}. Plotted per-timepoint sample means of the free energy under both flow regimes are contained in Figure \ref{fig:mean_F}. By adding the precision-weighted prediction error (KL) term to the internal updates, $F$ has a faster decrease and a greater magnitude of decrease under \eqref{example-eq-iii}; this is evidently from the prediction error driving the flow of $\partial_\mu F$ to align $q(\eta; \mu)$ with $p(\eta \mid s, a)$.

\begin{figure}[htb!]
    \centering
    \includesvg{internal.svg}
    \caption{{\bf Evolution of the internal state under $\lag$ and under $F$.} The dynamics of internal states in two regimes, plotted on a polar spiral showing the evolution in the phase in time. The dynamics under a gradient descent on $\lag$ and on $F$ show excellent agreement over all timescales.}
    \label{fig:internal}
\end{figure}

\begin{figure}[htb!]
    \centering
    \includesvg{a_tr.svg}
    \caption{{\bf Evolution of the active state under $\lag$ and under $F$.} The dynamics of the active state in two regimes were also plotted; again, excellent agreement is observed.}
    \label{fig:actions}
\end{figure}

\begin{figure}[htb!]
    \centering
    \includesvg[scale=0.8]{mean_KL_F.svg}
    \caption{{\bf Decrease of the KL divergence under the flow on $F$.} The sample mean ($N = 200$) of the KL divergence along trajectories of \eqref{example-eq-iii} is shown.}
    \label{fig:KL}
\end{figure}

\begin{figure}[htb!]
    \centering
    \includesvg{mean_F.svg}
    \caption{{\bf Decrease of the free energy under trajectories in both flow regimes.} The sample mean ($N = 200$) of the variational free energy along trajectories of \eqref{example-eq-ii} and along trajectories of \eqref{example-eq-iii} are compared. Both drive the free energy towards zero, however, the flow under $F$ decreases more rapidly and attains a lesser minimum closer to zero.}
    \label{fig:mean_F}
\end{figure}

The reason an inferential interpretation is useful here is in the coding of stimuli in neurones and neuronal ensembles. Minimising variational free energy, and the resultant model of internal and active states behaving as if they were performing Bayesian inference, endows the phase of the oscillation with the interpretation of a posterior belief about expected incoming signals in the environment and the Kalman-like updates as modulations of gain to minimise precision-weighted prediction error, rather than merely a position on a limit cycle. Under the relationship to stochastic thermodynamics forged in {\it e.g.} \cite{parr2020markov}, quantitative predictions can be made about the cost of maintaining such beliefs---and with an explicit low-dimensional representation of such signals given by the concentration of the non-equilibrium steady state density, we have an organically emerging neural manifold from those beliefs. 

\subsection{Looking ahead}

It is important to consider that we have not quite taken an agnostic approach, {\it i.e.}, taken an existing model, extracted a free energy gradient from it, and then tested our principle on it. We begin with a generative model and a choice of non-equilibrium steady state, and then engineer a Markov-blanketted Langevin dynamics of the form \eqref{sd-eq} with prescribed $Q$ and $\Gamma$, showing in detail how these dynamics can be written as a flow on variational free energy. Since the periodically firing cells model above has been crafted in a bespoke fashion to have a prescribed $Q$, $\Gamma$, and non-equilibrium steady state, it would be interesting to contrast this with a given set of equations, validated by independent modelling work, and treat it using the above reasoning. That is to say, the example was explicitly designed so that the solenoidal term $Q$ and dissipative term $\Gamma$ are convenient (and, for simplicity, taken to be constant), making the inferential structure and thermodynamic quantities analytically transparent. This means that the example is not an independent empirical test of the FEP, but a carefully crafted illustration of what the formalism looks like when the required decomposition is assumed to exist. However---this top-down approach illustrates how our FEP might be used to model a system and explain some observed behaviours inferentially, once a dynamical system built around those behaviours is given. The examples in \S3.1--3.2 are explicit constructions that exploit this fact: starting from a chosen non-equilibrium steady state and generative model, they show how the FEP provides a tractable representation of self-organisation (to a fixed point or limit cycle) as approximate Bayesian inference. A more demanding, `bottom-up' demonstration would begin from an independently specified stochastic dynamical model---for example, a physically motivated Langevin equation---and then: (i) derive its steady-state density $p^*(x)$ and probability current, (ii) determine whether there exist fields $Q(x)$ and $\Gamma(x)$ satisfying the constraints of \eqref{sd-eq}, and if so, (iii) read off the corresponding free energy functional and variational interpretation. This is precisely the kind of agnostic derivation that would address worries about circularity: instead of constructing a system to fit a free-energy-minimising flow, one would show that a pre-existing model can be recast in this way under clearly stated assumptions.

Our central contribution in the present paper is therefore more modest but, we would argue, still substantive. Namely, we show that, whenever a decomposition such as \eqref{sd-eq} and a Markov blanket exist, one can construct an associated family of generative models and variational densities such that the physical dynamics can be written as descending a variational free energy functional. On this basis, the two toy systems are offered as worked examples of this constructive use of the FEP, rather than as empirical validations of its universality.

\section{Some preliminary philosophical considerations}\label{prelim-phil}

We will now enter a more philosophical discussion concerning the {\it use} of the FEP, guiding its application and interpretation. The preceding technical arguments establish the FEP as an explanatory principle that can be applied to any thing (particle, person, or population) around which one can draw a boundary or Markov blanket. This means that the FEP is what it says on the tin: it is a principle. Like the principles of least action or of maximum entropy, the FEP is a piece of machinery that produces models \cite{ramstead2022bayesian}. 

Throughout, `as if' marks an interpretive equivalence: internal flows that minimise variational free energy can be read as variational inference without presupposing explicit, propositional models inside the system. Where specific mechanisms are identified ({\it e.g.}, precision-weighted message passing), this reading coincides with literal implementation; otherwise, it remains an explanatory stance. How might one apply the FEP to make a model of a particular particle, person or population? The answer is straightforward: the observer is trying to infer the generative model\textemdash which contains unobservable internal states of an object\textemdash that best explains the observed dynamics of some thing. What is observed is simply the blanket states of some `thing' that shield internal states from direct observation. If a generative model that best explains an object's behaviour can be identified, then by construction the FEP affords the modeller a complete (but not over-complete) description of the object's internal dynamics, even though they can never be observed. As noted above, the technical term for this relationship between real-world system and generative model is ``entailment'': a system that conforms to the dependence structure of the generative model is said to {\it entail} that generative model, with non-entailment of a known generative model being surprising.

If we now move to a meta-theoretical perspective and consider one particle ({\it e.g.}, a scientist or philosopher) observing another particle, we have the interesting situation where the modeller may impute a particular generative model that best explains the mechanics of the observed particle. Is this meta-model a map or a territory?\footnote{This meta-theoretical move is commonplace in practical applications of the FEP and is sometimes referred to as meta-Bayesian inference or, more simply, observing the observer \cite{daunizeau}. Practically, it involves optimising the parameters of a generative model, such that under the FEP (the ideal Bayesian observer assumption) the observable behaviour of the particle is rendered the most likely. This application of the FEP is often described as computational phenotyping \cite{schwartenbeck}.} Using a generative model to capture the formal structure of the environment\textemdash and to distinguish it from the model that is carried by particles (the variational density)\textemdash can be read as a new take on the \emph{nouvelle AI} idea that physical systems are their own best models; or as a new take on the good regulator theorem. For instance, in work on robotics following the tradition of embodied cognition \cite{beer1995computational, brooks1991intelligence}, practitioners have eschewed the construction of agents with internal representations of some environment. In these systems, the physics and geometry of the situation were sufficient to endow these agents with the capacity to couple to an environment. The world is, on this view, its own best representation. In our view, this approach dovetails nicely with FEP-theoretic modelling, where the generative model in FEP-theoretic constructions is just a joint probability density defined over all the states (or paths) of a system. In other words, the generative model is just our representation of the world as we believe it to be; and this model need not be encoded directly in the particle or agent. Given a generative model or Lagrangian of the appropriate sort, we can show that the system evinces a particular partition ({\it i.e.}, contains particles); and we can show that subsets of the system track each other, where tracking means inferring or becoming the sufficient statistics of probabilistic beliefs about their external states. We interpret this tracking as a form of inference, namely, variational or approximate Bayesian inference under a generative model ({\it i.e.}, the map). 

In the context of this meta-question, one may ask how an observer could come to consciously think an observed system is manipulating a model if the observer's own brain is simply flowing down a free-energy gradient. Here, an observer's conscious report that `X is using an internal model' may arise from some higher-order generative model over another agent's (or one's own) internal states, {\it i.e.}, beliefs about the other system's blanket-to-internal mappings (what we have called `observing the observer'). Conscious thought in this context is a higher-level subset of those internal states which summarises beliefs about others (and the self) within a hierarchical model. Mechanistically, this report is implemented by the observer's own free-energy-minimising dynamics. 

In short, if we equate the territory with the full joint system (crucially, including the unobserved internal states or dynamics of the observed thing), then the generative model plays the role of a map that best describes that (partially observed) territory. Note here, that the map only exists in relation to the observer\textemdash and, crucially, that the observer could be observing itself. In addition, the application of the FEP enables one to map out the parts of a territory that are unobservable in another sense; namely, in the sense that one cannot observe every state a given thing has been, or will be in. Even if one could, the internal states would be forever sequestered behind blanket states. Applications of the FEP place the generative model centre-stage as explanations for the behaviour of things\textemdash in terms of a map of some territory that can only be observed through its impressions on its Markov blanket. Here both observed and observer are described at the mechanistic level by flows on free energy gradients; at the informational level, the observer's higher-order model licenses attributions like `the observed system is manipulating an internal model'.

\section{A map of that part of the territory that behaves as if it were a map}\label{map-of-maps-sec}

Interestingly, because it incorporates this information about things that are coupled and their coupling, it has been suggested that our FEP conflates the metaphorical map (our mathematical model of the thing modelled) and the territory (the thing modelled). The so-called \textit{model reification} or \textit{map-territory} fallacy is a general critique of the practice of scientific modelling\textemdash in physics, principally due to Cartwright \cite{cartwright1984laws, cartwright1989nature}. Some have referenced this fallacy in relation to prior work on the FEP \cite{Bruineberg2020, VanEs2020}. In particular, it has been suggested that FEP-theoretic modelling conflates the metaphorical ``map''\textemdash ({\it i.e.}, the scientific model that scientists use to make sense of some real-world phenomenon)\textemdash and ``territory,'' {\it i.e.}, the actual physical system that is being modelled. (See \cite{Andrews2021, andrews2022making, kirchhoff2022literalist, ramstead2022bayesian} for related critical discussions of such claims.) According to this line of thinking, using the FEP to claim that we can model objects as themselves engaging in inference about the statistics of their environment constitutes a case of model reification. The allegation is that, in describing the dynamics of physical objects as implementing a form of inference\textemdash as opposed to considering the inferential aspects of the explanation as pertaining to our {\it scientific models} of those objects\textemdash the FEP theorist mistakenly conflates their metaphorical ``map'' of the territory ({\it i.e.}, the scientific, FEP-theoretic model) with the ``territory'' (the real-world system itself). This constitutes a reification of our model by assuming some aspect of the model or ``map'' is a real feature of the physical world or ``territory'' \cite{Andrews2021, andrews2022making}. By clarifying that this potential difficulty does not directly and necessarily apply to FEP-theoretic modelling, we are able to turn the story on its head: namely, we point out that the FEP provides a set of ultimate explanatory constraints on what counts as a model (or map) of any {\it thing} that exists in the physical world.

We argue that the FEP can be understood, metaphorically, as a ``map'' of sorts: a map of that part of the territory which behaves or looks as if it were a map. In other words, the FEP provides us with tools to understand the mathematically striking property of self-organising systems: that they look as if they infer, track, reflect, or represent the statistics of their environment. Even more simply, it gives us tools to model systems that look as if they are modelling the world.

How does FEP-theoretic technology allow us to model self-organising systems as modelling their world? The generative model or joint probability $p(\eta, b, \mu)$ plays a central role. This generative model can be read as a probabilistic specification of the states or paths characteristically occupied by the joint particle-environment system. In other words, our scientific model (or map) includes the generative model entailed by the system ({\it i.e.}, the territory). In this sense, the generative model we hypothesise\textemdash to account for observations of an object\textemdash allows us to represent the constraints on the kinds of states the joint system can be found in. In other words, the generative model is a modeller's map of the whole ``territory'' that characterises the observed system and its environment (including the modeller, as an observer in the environment). 

The technology of the FEP allows us to say that certain components of the system being considered look as if they track other components, and that this is a feature of our map contained in a generative model under a particular partition. In providing us with a calculus allowing us to model things as modelling other things, the FEP provides us with a map (a scientific model) of any possible map or tracking relationship between things (namely, a mapping between internal and external states, mediated by blanket states).\footnote{Here, we are assuming representationalist accounts of scientific practice. For alternatives, see \cite{Andrews2021}.}

Does our ability to always describe some dynamics as a form of inference imply that those dynamics are literally a form of inference, as opposed to being a manner of describing those dynamics? Here, it is important to defuse a possible (but in our view na\"ive) objection. It is a truism that physical systems need not explicitly calculate their trajectories of motion, to be modelled as pursuing such trajectories. In developing an inferential account of the dynamics of systems, we need not assume that the particle itself is literally performing inference.\footnote{Ways of defending stronger, increasingly literal versions of this positions are available as well. For a defence of the claim that physical systems are quite literally in the business of inference, see \cite{Kiefer2017literal, Kiefer2020psycho}. Related to this, views of physical systems implementing computations has been articulated \cite{horsman2014does, aaronson2005guest}, as well as views where systems self-organise by computing predictions \cite{still2012thermodynamics, perunov2016statistical, ueltzhoffer2021drive, seifert2019stochastic}.} What is at stake with the FEP is an `as if' description. We simply take the fact that surprisal varies with the dynamics of the system, which implies free energy is minimised; the minimisation of this quantity is mathematically equivalent to inference, in the sense of being an estimator.

In summary, our scientific, FEP-theoretic model is, metaphorically, a map that allows us to say that some (internal) subset of the system looks as if it possesses a map; which we interpret formally as tracking the statistics of another (external) subset. \textit{Our} map (as scientists and modellers) can be identified in an unproblematic way as the generative model of the system, which FEP-theoretic technology enables us to write down. It is by construction that our model is a model of the \textit{modelling capacities of subsets of the real-world system}: that is, our map is precisely a map of how objects behave as if they were maps, allowing us to construct scientific models of the maps encoded (or looking as if they were encoded) by the internal subset of the system considered. This is an answer to allegations that FEP-theoretic modelling commits the map-territory fallacy: namely, it shows that there is no conflation of the map and the territory, in the sense of the map we {\it postulate} as being embodied by the system within the mathematical machinery of the FEP. We remark that we do \textit{not} claim to invalidate the map-territory fallacy itself. That is, we do not suggest that the map and the territory are never conflated; they obviously are in cases of genuine model reification (as discussed cogently by \cite{andrews2022making}). Models built using the FEP are still scientific models, and the properties of FEP-theoretic models also run the risk of reification. Instead, we are addressing the claim that the FEP is limited in its effectiveness or engages in model reification because real physical things cannot be modelled as themselves modelling their environments by minimising free energy.

Taking all of the above, we might say that the FEP ultimately provides us with a map of \textit{any possible map whatsoever} of, or held by, a physical system. This echoes seminal work in philosophy by Wittgenstein \cite{wittgenstein2013tractatus}. In his famous \textit{Tractatus Logico-Philosophicus}, Wittgenstein set out to delimit the domain of sensible propositions by determining the general form of a proposition, an utterance able to carry a truth value. This set ultimate constraints on what is sensible, and bounded what is meaningful ``from inside'' of language itself. Similarly, the FEP provides ultimate constraints on possible maps or modelling relations, which arise from what it means to be a map or model of a physical process at all. Indeed, any modelling process consistent with the laws of physics must conform to the FEP, such that the FEP sets ultimate limits on what can count as a map or model\textemdash starting from within the technology of mapping or modelling itself.

In sum, the generative model is a kind of scientific model, which harnesses what we know about the dependency structure of the system. A variational density, on the other hand, is a probability density over external states or paths, implemented in terms of the sufficient statistics of external states given blanket states. It is an explicitly inferential representation of the manner in which subsets of a system with a particular partition track each other: in which case, we are describing a part of the territory that, as seen in our map, also behaves as if it were a map.

\section{Conclusion}

We argued that, in describing `things' or `particles' as estimators or (deflated) representations of the systems to which they are coupled, the FEP-theoretic apparatus does not commit the map territory fallacy, {\it i.e.}, it is false that applications of the FEP necessarily reify aspects of the metaphorical map ({\it i.e.}, our scientific model), mistakenly taking them to be part of the territory ({\it i.e.}, the real-world system that we wish to model); although this remains possible in principle. We have further argued that this allegation itself constitutes a \textit{map-territory fallacy fallacy}. In distinguishing the generative model and the variational density\textemdash and the generative model and the generative process from the real-world system that we aim to model\textemdash the technology of FEP-theoretic modelling allows us to construct a map of that part of the territory which \textit{looks to an external observer as if} or \textit{behaves as if it were a map}, without committing model reification. The FEP is thus, metaphorically, a map of any possible map whatsoever of, or held by, a physical system. One ought to celebrate this `territory-map mapping,' and the FEP is ultimately, at its core, a principled approach to the formalisation of this mapping.

\bibliographystyle{unsrt}
\bibliography{main}

@article{DaCosta2021,
    title = {{Bayesian mechanics for stationary processes}},
    author = {{Da Costa}, Lancelot and Friston, Karl J and Heins, Conor and Pavliotis, Grigorios A},
    journal = {Proceedings of the Royal Society A},
    year = {2021},
    number = {2256},
    volume = {477},
    doi = {10.1098/rspa.2021.0518}
}

@article{ramstead2022bayesian,
  title={On {B}ayesian mechanics: a physics of and by beliefs},
  author={Ramstead, Maxwell J D and Sakthivadivel, Dalton A R and Heins, Conor and Koudahl, Magnus and Millidge, Beren and Da Costa, Lancelot and Klein, Brennan and Friston, Karl J},
  journal={Interface Focus},
  volume={13},
  number={3},
  pages={20220029},
  year={2023},
  publisher={The Royal Society}
}

@article{Sakthivadivel2022b,
    title = {{Towards a geometry and analysis for Bayesian mechanics}},
    author = {Sakthivadivel, Dalton A R},
    year = {2022},
    note = {Preprint arXiv:2204.11900},
}

@inproceedings{classical-physics,
  title={A worked example of the {B}ayesian mechanics of classical objects},
  author={Sakthivadivel, Dalton A R},
  booktitle={Active Inference: Third International Workshop},
  volume={1721},
  series={Communications in Computer and Information Science},
  pages={298--318},
  year={2023},
  publisher={Springer}
}

@article{sakthivadivel2022d,
  title={Weak {M}arkov Blankets in High-Dimensional, Sparsely-Coupled Random Dynamical Systems},
  author={Sakthivadivel, Dalton A R},
  note={Preprint arXiv:2207.07620},
  year={2022}
}

@article{Bruineberg2020,
   author = {Bruineberg, Jelle and Dolega, Krzysztof and Dewhurst, Joe and Baltieri, Manuel},
   title = {The Emperor’s New {M}arkov Blankets},
   journal={Behavioral and Brain Sciences},
   pages={1--63},
   year={2020},
   publisher={Cambridge University Press}
}

@article{VanEs2020,
   author = {van Es, Thomas},
   title = {Living models or life modelled? {O}n the use of models in the free energy principle},
   journal = {Adaptive Behavior},
   ISSN = {1059-7123},
   year = {2020},
   type = {Journal Article}
}

@article{andrews2022making, 
title={Making reification concrete: a response to {B}ruineberg et al.}, 
volume={45}, 
DOI={10.1017/S0140525X22000310}, 
journal={Behavioral and Brain Sciences}, 
publisher={Cambridge University Press}, 
author={Andrews, Mel}, year={2022}, 
pages={e186}
}

@article{brooks1991intelligence,
  title={Intelligence without representation},
  author={Brooks, Rodney A},
  journal={Artificial Intelligence},
  volume={47},
  number={1-3},
  pages={139--159},
  year={1991},
  publisher={Elsevier}
}

@incollection{beer1995computational,
  title={Computational and dynamical languages for autonomous agents},
  author={Beer, Randall D},
  booktitle={Mind as Motion: Explorations in the Dynamics of Cognition},
  pages={121--147},
  year={1996}
}

@article{kirchhoff2022literalist,
  title={The Literalist Fallacy and the Free Energy Principle: model-Building, Scientific Realism, and Instrumentalism},
  author={Kirchhoff, Michael D and Kiverstein, Julian and Robertson, Ian},
  journal={The British Journal for the Philosophy of Science},
  year={2022}
}

@article{Ramstead2020semantics,
   author = {Ramstead, Maxwell J D and Friston, Karl J and Hip\'olito, In\^es},
   title = {Is the free-energy principle a formal theory of semantics? {F}rom variational density dynamics to neural and phenotypic representations},
   journal = {Entropy},
   year = {2020},
   type = {Journal Article}
}

@article{Kiefer2020psycho,
  title={Psychophysical identity and free energy},
  author={Kiefer, Alex B},
  journal={Journal of the Royal Society Interface},
  volume={17},
  number={169},
  pages={20200370},
  year={2020},
  publisher={The Royal Society}
}

@article{Ramstead2019enactive,
   author = {Ramstead, Maxwell J D and Kirchhoff, Michael D and Friston, Karl J},
   title = {A tale of two densities: active inference is enactive inference},
   journal = {Adaptive Behavior},
   year = {2020},
   volume={28},
   number={4},
   pages={225--239},
}

@article{Friston2019,
    title = {{A free energy principle for a particular physics}},
    author = {Friston, Karl J},
    year = {2019},
    note = {Preprint arXiv:1906.10184}
}

@inproceedings{Kiefer2017literal,
  title={Literal perceptual inference},
  author={Kiefer, Alex B},
  booktitle={Philosophy and Predictive Processing},
  year={2017}
}

@article{horsman2014does,
  title={When does a physical system compute?},
  author={Horsman, Clare and Stepney, Susan and Wagner, Rob C and Kendon, Viv},
  journal={Proceedings of the Royal Society A},
  volume={470},
  number={2169},
  pages={20140182},
  year={2014},
  publisher={The Royal Society Publishing}
}

@article{Andrews2021,
    title = {{The math is not the territory: navigating the free energy principle}},
    author = {Andrews, Mel},
    journal = {Biology \& Philosophy},
    year = {2021},
    volume = {36},
    number = {3},
    pages = {30},
    doi = {10.1007/s10539-021-09807-0}
}

@book{wittgenstein2013tractatus,
  title={{Tractatus Logico-Philosophicus}},
  author={Wittgenstein, Ludwig},
  year={1922},
  note={2013 edition, Routledge}
}

@article{jaynes1,
  title={Information theory and statistical mechanics},
  author={Jaynes, Edwin T},
  journal={Physical Review},
  volume={106},
  number={4},
  pages={620},
  year={1957},
  publisher={APS}
}

@book{jaynes2,
  title={Probability Theory: The Logic of Science},
  author={Jaynes, Edwin T},
  year={2003},
  publisher={Cambridge University Press}
}

@article{daunizeau,
  title={Observing the observer ({I}): meta-{B}ayesian models of learning and decision-making},
  author={Daunizeau, Jean and Den Ouden, Hanneke E M and Pessiglione, Matthias and Kiebel, Stefan J and Stephan, Klaas E and Friston, Karl J},
  journal={PLoS ONE},
  volume={5},
  number={12},
  pages={e15554},
  year={2010},
  publisher={Public Library of Science San Francisco, USA}
}

@article{schwartenbeck,
  title={Computational phenotyping in psychiatry: a worked example},
  author={Schwartenbeck, Philipp and Friston, Karl J},
  journal={eNeuro},
  volume={3},
  number={4},
  year={2016},
  publisher={Society for Neuroscience}
}

@book{cartwright1984laws,
  title={How the Laws of Physics Lie},
  author={Cartwright, Nancy},
  year={1984},
  publisher={Oxford University Press}
}

@book{cartwright1989nature,
  title={Nature's Capacities and their Measurement},
  author={Cartwright, Nancy},
  year={1989},
  publisher={Oxford University Press}
}

@phdthesis{beal2003variational,
  title={Variational algorithms for approximate Bayesian inference},
  author={Beal, Matthew J},
  year={2003},
  school={University of London, University College London}
}

@article{aaronson2005guest,
  title={Guest column: {NP}-complete problems and physical reality},
  author={Aaronson, Scott},
  journal={ACM SIGACT News},
  volume={36},
  number={1},
  pages={30--52},
  year={2005},
  publisher={ACM New York, NY, USA}
}

@article{still2012thermodynamics,
  title={Thermodynamics of prediction},
  author={Still, Susanne and Sivak, David A and Bell, Anthony J and Crooks, Gavin E},
  journal={Physical Review Letters},
  volume={109},
  number={12},
  pages={120604},
  year={2012},
  publisher={APS}
}

@article{perunov2016statistical,
  title={Statistical physics of adaptation},
  author={Perunov, Nikolay and Marsland, Robert A and England, Jeremy L},
  journal={Physical Review X},
  volume={6},
  number={2},
  pages={021036},
  year={2016},
  publisher={APS}
}

@article{seifert2019stochastic,
  title={From stochastic thermodynamics to thermodynamic inference},
  author={Seifert, Udo},
  journal={Annual Review of Condensed Matter Physics},
  volume={10},
  pages={171--192},
  year={2019},
  publisher={Annual Reviews}
}

@article{ueltzhoffer2021drive,
  title={A drive towards thermodynamic efficiency for dissipative structures in chemical reaction networks},
  author={Ueltzh{\"o}ffer, Kai and Da Costa, Lancelot and Cialfi, Daniela and Friston, Karl J},
  journal={Entropy},
  volume={23},
  number={9},
  pages={1115},
  year={2021},
  publisher={MDPI}
}

@article{friston2023path,
  title={Path integrals, particular kinds, and strange things},
  author={Friston, Karl and Da Costa, Lancelot and Sakthivadivel, Dalton A R and Heins, Conor and Pavliotis, Grigorios A and Ramstead, Maxwell and Parr, Thomas},
  journal={Physics of Life Reviews},
  volume={47},
  pages={35--62},
  year={2023},
  publisher={Elsevier}
}

@article{parr2020markov,
  title={Markov blankets, information geometry and stochastic thermodynamics},
  author={Parr, Thomas and Da Costa, Lancelot and Friston, Karl},
  journal={Philosophical Transactions of the Royal Society A},
  volume={378},
  number={2164},
  pages={20190159},
  year={2020},
  publisher={The Royal Society Publishing}
}

@Inbook{Pearl1998,
author="Pearl, Judea",
title={Graphical Models for Probabilistic and Causal Reasoning},
bookTitle="Quantified Representation of Uncertainty and Imprecision",
year="1998",
publisher="Springer",
pages="367--389"
}

@article{seifert2012stochastic,
  title={Stochastic thermodynamics, fluctuation theorems and molecular machines},
  author={Seifert, Udo},
  journal={Reports on Progress in Physics},
  volume={75},
  number={12},
  pages={126001},
  year={2012},
  publisher={IOP Publishing}
}

@article{ao2004potential,
  title={Potential in stochastic differential equations: novel construction},
  author={Ao, Ping},
  journal={Journal of Physics A: Mathematical and General},
  volume={37},
  number={3},
  pages={L25},
  year={2004},
  publisher={IOP Publishing}
}

@article{da2023entropy,
  title={The entropy production of stationary diffusions},
  author={Da Costa, Lancelot and Pavliotis, Grigorios A},
  journal={Journal of Physics A: Mathematical and Theoretical},
  volume={56},
  number={36},
  pages={365001},
  year={2023},
  publisher={IOP Publishing}
}

@article{mtff,
  title={On the map-territory fallacy fallacy},
  author={Ramstead, Maxwell J D and Sakthivadivel, Dalton A R and Friston, Karl J},
  note={Preprint arXiv:2208.06924},
  year={2022}
}

@article{friston2015knowing,
  title={Knowing one's place: a free-energy approach to pattern regulation},
  author={Friston, Karl and Levin, Michael and Sengupta, Biswa and Pezzulo, Giovanni},
  journal={Journal of the Royal Society Interface},
  volume={12},
  number={105},
  pages={20141383},
  year={2015},
  publisher={The Royal Society}
}

@book{freidlin_wentzell_2012,
  author    = {Mark I Freidlin and Alexander D Wentzell},
  title     = {Random Perturbations of Dynamical Systems},
  series    = {Grundlehren der mathematischen Wissenschaften},
  volume    = {260},
  year      = {1998},
  publisher = {Springer}
}

@article{friston2007variational,
  title={Variational free energy and the {L}aplace approximation},
  author={Friston, Karl and Mattout, J{\'e}r{\'e}mie and Trujillo-Barreto, Nelson and Ashburner, John and Penny, Will},
  journal={NeuroImage},
  volume={34},
  number={1},
  pages={220--234},
  year={2007},
  publisher={Elsevier}
}

@inproceedings{millidge2020relationship,
  title={On the relationship between active inference and control as inference},
  author={Millidge, Beren and Tschantz, Alexander and Seth, Anil K and Buckley, Christopher L},
  booktitle={Active Inference: First International Workshop},
  pages={3--11},
  year={2020},
  organization={Springer}
}

@article{pezzulo2018hierarchical,
  title={Hierarchical active inference: a theory of motivated control},
  author={Pezzulo, Giovanni and Rigoli, Francesco and Friston, Karl J},
  journal={Trends in Cognitive Sciences},
  volume={22},
  number={4},
  pages={294--306},
  year={2018},
  publisher={Elsevier}
}

@article{friston2012active,
  title={Active inference and agency: optimal control without cost functions},
  author={Friston, Karl and Samothrakis, Spyridon and Montague, Read},
  journal={Biological Cybernetics},
  volume={106},
  pages={523--541},
  year={2012},
  publisher={Springer}
}

@inproceedings{baioumy2021active,
  title={Active inference for integrated state-estimation, control, and learning},
  author={Baioumy, Mohamed and Duckworth, Paul and Lacerda, Bruno and Hawes, Nick},
  booktitle={2021 IEEE International Conference on Robotics and Automation},
  pages={4665--4671},
  year={2021},
  organization={IEEE}
}

@article{pio2016active,
  title={Active inference and robot control: a case study},
  author={Pio-Lopez, L{\'e}o and Nizard, Ange and Friston, Karl and Pezzulo, Giovanni},
  journal={Journal of The Royal Society Interface},
  volume={13},
  number={122},
  pages={20160616},
  year={2016},
  publisher={The Royal Society}
}

@article{pezzulo2015active,
  title={Active inference, homeostatic regulation and adaptive behavioural control},
  author={Pezzulo, Giovanni and Rigoli, Francesco and Friston, Karl},
  journal={Progress in Neurobiology},
  volume={134},
  pages={17--35},
  year={2015},
  publisher={Elsevier}
}

@article{baltieri2019pid,
  title={{PID} control as a process of active inference with linear generative models},
  author={Baltieri, Manuel and Buckley, Christopher L},
  journal={Entropy},
  volume={21},
  number={3},
  pages={257},
  year={2019},
  publisher={MDPI}
}

@article{friston2010generalised,
  title={Generalised filtering},
  author={Friston, Karl and Stephan, Klaas and Li, Baojuan and Daunizeau, Jean},
  journal={Mathematical Problems in Engineering},
  volume={2010},
  number={1},
  pages={621670},
  year={2010},
  publisher={Wiley Online Library}
}

@article{yuan2017sde,
  title={{SDE} decomposition and {A}-type stochastic interpretation in nonequilibrium processes},
  author={Yuan, Ruoshi and Tang, Ying and Ao, Ping},
  journal={Frontiers of Physics},
  volume={12},
  pages={1--9},
  year={2017},
  publisher={Springer}
}

@inproceedings{ma2015,
 author = {Ma, Yi-An and Chen, Tianqi and Fox, Emily},
 booktitle = {Advances in Neural Information Processing Systems},
 pages = {2917--2925},
 title = {A Complete Recipe for Stochastic Gradient {MCMC}},
 volume = {28},
 year = {2015}
}

\end{document}